# Error-speed correlations in biopolymer synthesis

Davide Chiuchiú,[1] Yuhai Tu,[2] and Simone Pigolotti[1, *]

[1]*Biological Complexity Unit, Okinawa Institute of Science and Technology Graduate University, Onna, Okinawa 904-0495, Japan*
[2]*IBM T.J. Watson Research Center, Yorktown Heights, NY 10598, U.S.A.*

Synthesis of biopolymers such as DNA, RNA, and proteins are biophysical processes aided by enzymes. Performance of these enzymes is usually characterized in terms of their average error rate and speed. However, because of thermal fluctuations in these single-molecule processes, both error and speed are inherently stochastic quantities. In this paper, we study fluctuations of error and speed in biopolymer synthesis and show that they are in general correlated. This means that, under equal conditions, polymers that are synthesized faster due to a fluctuation tend to have either better or worse errors than the average. The error-correction mechanism implemented by the enzyme determines which of the two cases holds. For example, discrimination in the forward reaction rates tends to grant smaller errors to polymers with faster synthesis. The opposite occurs for discrimination in monomer rejection rates. Our results provide an experimentally feasible way to identify error-correction mechanisms by measuring the error-speed correlations.

Organisms encode genetic information in heteropolymers such as DNA and RNA. Replication of these heteropolymers is a non-equilibrium process catalyzed by enzymes. The crucial observables to characterize these enzymes are their error rate and speed. A low error, defined as the fraction of monomers in the copy that do not match the template, ensures correct trasmission of biological information. High processing speed is also crucial to guarantee fast cell growth. Theoretical approaches have been developed to compute the average error and average speed of polymerization processes [1–7]. However, at the single molecule level, both error and speed can present significant stochastic fluctuations.

In this Letter we address fluctuations in the error and speed of polymer synthesis. In particular, we show that correlations between these quantities exist. These correlations provide a way to identify the error correction mechanism adopted by an enzyme from experimental data. This approach can circumvent the characterization of these enzymes by measuring all kinetic rates of the underlying reaction network [8–15].

We consider an enzyme that replicates an existing template polymer by sequentially incorporating monomers into a copy polymer (Figure 1a). In a given time interval $T$, the enzyme synthesizes a copy made up of a number of monomers $L$. Because of thermal fluctuations, enzymes sometimes incorporate wrong monomers ($w$) that do not match the template, instead of the right ones ($r$). In practical cases, there can be multiple types of wrong monomers; for simplicity, we do not distinguish among them. We denote $R$ as the number of right matches and $W$ the number of wrong matches in the copy, so that $R + W = L$. The error of the polymer copy can be then expressed as

$$\eta = \frac{W}{L}. \tag{1}$$

We focus on two possible setups, corresponding to two idealized experiments. In the first, the enzyme replicates a given template polymer for a fixed time $T \gg 1$. (Fig. 1b). Due to the stochasticity of single-molecule biochemical reactions, both the polymer length $L$ and the error $\eta$ fluctuate. We denote their variance with $\sigma_L^2 = \langle L^2 \rangle - \langle L \rangle^2$, $\sigma_\eta^2 = \langle \eta^2 \rangle - \langle \eta \rangle^2$ and the covariance with $\sigma_{\eta L}^2 = \langle \eta L \rangle - \langle \eta \rangle \langle L \rangle$, where $\langle \ldots \rangle$ is an average over different realizations of the same process. Since $T$ is fixed, we quantify the correlations between error and speed with the error-length coefficient

$$r_{\eta L} = \frac{\sigma_{\eta L}^2}{\sigma_L \sigma_\eta}. \tag{2}$$

In the second setup, each realization terminates when the enzyme has incorporated a number $L \gg 1$ of monomers (Fig. 1c). In this case, $L$ is fixed, whereas the total duration $T$ of the copy process fluctuates. This setup represents the biological scenario where an enzyme copies a polymer of fixed length. In this case, we study the correlation between the polymerization error and speed via the coefficient

$$r_{\eta T} = \frac{\sigma_{\eta T}^2}{\sigma_T \sigma_\eta} \tag{3}$$

where $\sigma_T^2 = \langle T^2 \rangle - \langle T \rangle^2$ is the variance of $T$ and $\sigma_{\eta T}^2 = \langle \eta T \rangle - \langle \eta \rangle \langle T \rangle$.

Our two setups are akin to two conjugate ensembles in equilibrium statistical physics. For large times (and lengths), fluctuations in these two ensembles can be related by means of large deviation theory [16]. Following this approach we obtain

$$r_{\eta T} = -r_{\eta L} \tag{4}$$

(see SI for details). Eq. (4) implies that the two setups correspond to two equivalent ensembles. Therefore, in the following we will focus on the fixed time setup only.

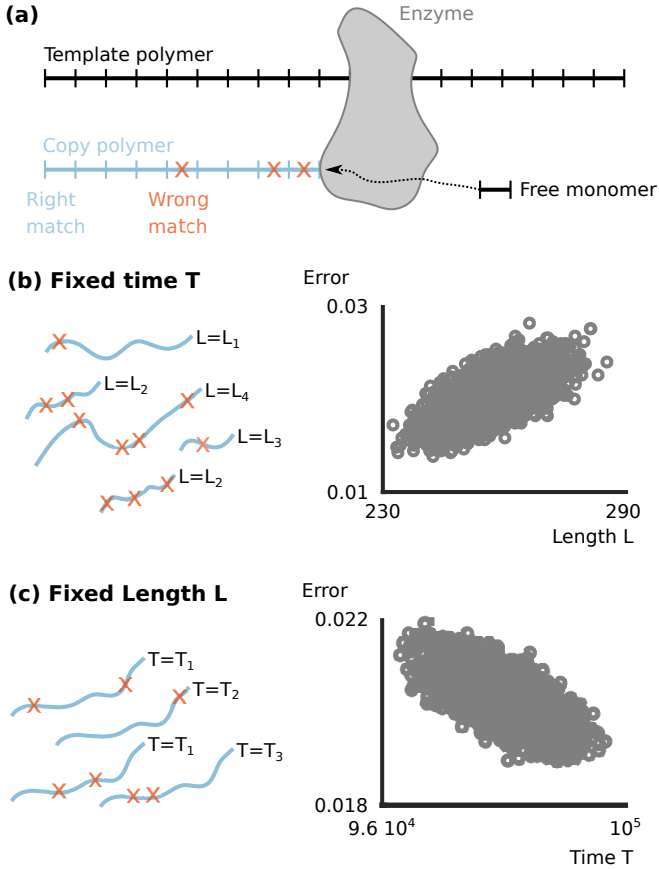

FIG. 1. (a) An enzyme reads an existing heteropolymer as a template and sequentially incorporates monomers to copy it. Each incorporated monomer can either be a right ($r$) or wrong ($w$) match with the template polymer. (b) Due to thermal fluctuations, the polymer length $L$ and error $\eta$ are random quantities at fixed completion time $T$. (c) When an enzyme produces a copy polymer with fixed length, the error $\eta$ and the time $T$ fluctuate. Scatterplots in (b) and (c) represent $N = 5000$ realizations of the same polymerization process where incorporation occurs via two sequential irreversible reactions, see SI. Data skewness indicates correlations in the observables.

To estimate $r_{\eta L}$ we first observe that the distributions of $R$ and $W$ tend to Gaussian for large $T$ due to the central limit theorem. We can therefore obtain the moments of $L = R + W$ and $\eta = W/(W + R)$ from those of $R$ and $W$. This procedure yields

$$r_{\eta L} = \frac{(1 - 2\langle\eta\rangle) \sigma_{RW}^2 + (1 - \langle\eta\rangle) \sigma_W^2 - \langle\eta\rangle \sigma_R^2}{\sqrt{\sigma_R^2 \sigma_W^2 - (\sigma_{RW}^2)^2}}. \quad (5)$$

To compute the quantities in Eq. (5), we assume that the final chemical reaction to incorporate a $r$ or a $w$ monomer is irreversible. This assumption is realistic for most practical cases such as DNA polymerization [10, 17] and protein translation [12–14]. Our framework could be generalized to cases where the last reaction is reversible, permitting an interpretation of the results using stochastic thermodynamics [1–5, 18]. For simplicity, we also assume that probabilities to incorporate right and wrong matches do not depend on the template monomer. Under these assumptions, we describe the polymerization process by means of the probabilities $\eta_0$ and $1 - \eta_0$ to incorporate a wrong ($w$) or a right ($r$) monomer, respectively, and the probability distributions $P(\tau|r)$ and $P(\tau|w)$ that it takes a time $\tau$ to incorporate an $r$ or a $w$ monomer, respectively. The value of $\eta_0$ and the functions $P(\tau|r)$ and $P(\tau|w)$ can be computed from the underlying reaction network [6, 19]. With these quantities we can express the joint probability $P(R, W|T)$ for large $T$ as

$$P(R, W|T) \approx \binom{R + W}{W} \eta_0^W (1 - \eta_0)^R \times \quad (6)$$

$$\times \int_0^\infty \prod_{i=1}^R \prod_{j=1}^W d\tau_i d\tau_j P(\tau_i|r) P(\tau_j|w) \delta\left(\sum_{n=1}^R \tau_n + \sum_{m=1}^W \tau_m - T\right).$$

In Eq. (6), the binomial term weight the probability of incorporating $R$ right and $W$ wrong monomers. The integral term in the second line selects trajectories whose sum of incorporation times is equal to $T$.

Evaluating the average error for large $T$ gives the consistency relation $\langle\eta\rangle = \eta_0$. Computing the covariance matrix of $P(R, W|T)$ in the same limit (see SI) and substituting the resulting moments in Eq. (5) gives

$$r_{\eta L} = \frac{\beta}{\sqrt{1 + \beta^2}} \quad (7)$$

with

$$\beta = \frac{(\langle\tau\rangle_r - \langle\tau\rangle_w) \sqrt{\eta_0 (1 - \eta_0)}}{\sqrt{(1 - \eta_0) \sigma_{\tau,r}^2 + \eta_0 \sigma_{\tau,w}^2}}, \quad (8)$$

where $\langle\tau\rangle_r$, $\langle\tau\rangle_w$, $\sigma_{\tau,r}^2$ and $\sigma_{\tau,w}^2$ are the means and variances of $P(\tau|r)$ and $P(\tau|w)$, that we assume to be finite. We validated Eqs. (7) and (8) with stochastic simulations (see SI) and we will use them to compute error-speed correlations in the following. Expanding Eq. (8) and Eq. (7) for small $\eta_0$ leads to

$$r_{\eta L} \approx \frac{\langle\tau\rangle_r - \langle\tau\rangle_w}{\sigma_{\tau,r}} \sqrt{\eta_0}. \quad (9)$$

Eq. (9) is our main result. It predicts that the sign of $r_{\eta L}$ depends on the sign of $(\langle\tau\rangle_r - \langle\tau\rangle_w)$ only. We will show that, in practice, the error correction mechanisms determine this sign.

*Kinetic Proofreading.* Hopfield's kinetic proofreading model [20] is an elegant example of an incorporation processes implementing error correction. In this model, the enzyme first captures either a $r$ or $w$ monomer (Figure 2.a). After the initial binding, the enzyme can either reject the monomer or consume ATP to induce a conformational change. Thanks to this change, the enzyme gains a second chance to reject wrong monomers. This second rejection reaction is the kinetic proofreading and it greatly reduce the error probability $\eta_0$. This idea has been generalized to more complex proofreading models [2, 6, 19, 21–24]. Rates of forward reactions in the Hopfield model do not depend on the monomer type, whereas rejection reactions have higher rates for $w$ than $r$ monomers (Figure 2.a). In the proofreading regime (Figure 2.a), the error probability $\eta_0$ can be estimated with first passage time techniques [19] as

$$\eta_0 \approx \left(1 + \frac{k_r}{k_w}\frac{k_p^r}{k_p^w}\right)^{-1} \approx e^{-\frac{2(\Delta E_w - \Delta E_r)}{k_B T}} \quad (10)$$

where the ratios $k_r/k_w$ and $k_p^r/k_p^w$ reflect the discrimination in the rejection rates (see [20] and SI), $k_b$ is the Boltzmann constant and $T$ is the temperature. Both these ratios relate to the difference $\Delta E_r - \Delta E_w$ in binding energy of $r$ and $w$ monomers through $k_r/k_w = k_p^r/k_p^w = \exp[(\Delta E_r - \Delta E_w)/k_B T]$. Outside of the error correction regime, the error is always larger than predicted by Eq. (10) [20, 25]. In the proofreading regime of the Hopfield model, error and speed fluctuations are positively correlated. In particular, the error-length coefficient always falls in the range

$$0 \le r_{\eta L} \le \eta_0 \left(\sqrt{\frac{1-\eta_0}{\eta_0}} - 1\right) \quad (11)$$

for any choice of $\eta_0$, see SI and Fig. 3. This implies that the error-speed correlations become negligible when proofreading ensures very small errors.

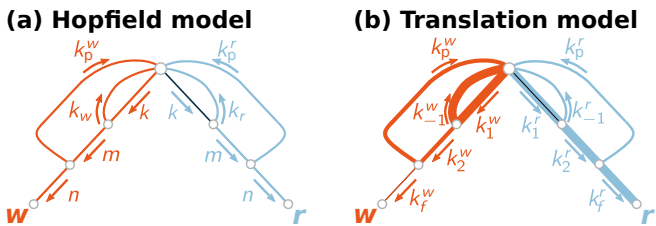
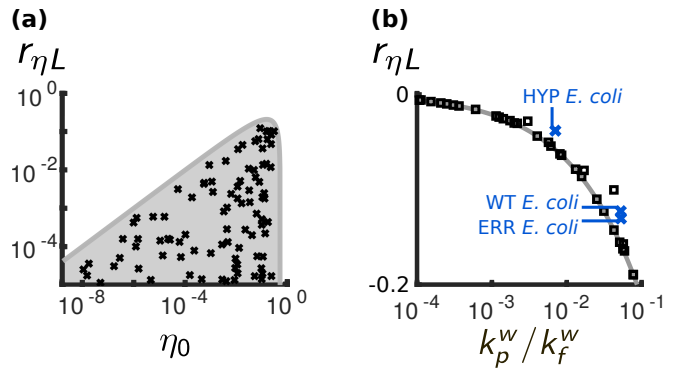

FIG. 2. Reaction networks for polymer synthesis. (a) Hopfield model. The kinetic rates satisfy the relations $k_r = k\exp[\Delta E_r/k_B T]$, $k_w = k\exp[\Delta E_w/k_B T]$, $k_p^r = m\exp[\Delta E_r/k_B T]$ and $k_p^w = m\exp[\Delta E_w/k_B T]$ with $k \gg 1$, $m = 1$ and $n \ll 1$, so that the model operates in the proofreading regime [20]. (b) Protein translation model from [19] with rates extracted from [14]. Same line thickness marks reaction rates of the same order of magnitude.

*Protein translation.* A standard model of protein translation is characterized by the same reactions of the Hopfield model (Figure 2.b and [19, 26, 27]). A major difference is that forward reactions discriminate between the $r$ and $w$ monomers (Table S1 and [14, 19]). Within this model we estimate the error probability as

$$\eta_0 \approx \frac{k_f^w}{k_f^r}\left(1 + \frac{k_f^w}{k_p^w}\right)^{-1} \quad (12)$$

FIG. 3. The Hopfield model and the protein translation model have opposite error-speed correlations. (a) Hopfield model. The gray shaded region defines the allowed values of $r_{\eta L}$ for a given error probability $\eta_0$, see Eq (11). Black crosses are estimates of $\eta_0$ and $r_{\eta L}$ for 60 random sets of reaction rates in the proofreading regime (see caption of Figure 2, SI, and Table S2). (b) Protein translation. To test Eq. (13) (gray line), we computed $r_{\eta L}$ with the kinetic rates in [19] for wild type *E. coli*, a hypercorrective and an error-prone mutation (blue crosses). We also evaluated $r_{\eta L}$ for randomly generated sets of the reaction rates in Figure 2.b (black squares). For all points in both panels, correlation coefficients are evaluated by means of Eqs. (7)-(8) upon computing moments of incorporation times with first passage time techniques [19]. See SI for details of numerical calculations.

(see [19] and SI). In this case, the error probability depends on the relative preference to bind $r$ rather than $w$ monomers (term $k_f^w/k_f^r$). Proofreading effectiveness over the incorporation reaction for $w$ monomers (term $k_f^w/k_p^w$) further decrease the error probability. Because of the discrimination in the forward rates, the energy difference $\Delta E_r - \Delta E_w$ does not set a lower bound to the error probability as in the Hopfield model [6]. Similar calculations as in Eq. (11) predict an error-length coefficient

$$r_{\eta L} \approx -\frac{1}{\sqrt{2}}\left(1 + \frac{k_p^w}{k_f^w}\right)^{-\frac{1}{2}}. \quad (13)$$

(SI and Figure 3). At variance with the Hopfield model, the error-length coefficient is always negative in protein translation. This striking difference arises from the discrimination in the forward rates, as further clarified in the following. Also in this case, the absolute value of the error-speed correlations decreases at increasing proofreading efficiency. Ribosomes with impaired kinetic proofreading should then exhibit stronger error-speed correlations. A computation of error-speed correlations from experimentally measured rates for different *E. coli* strains supports Eq. (13), Fig. 3.

*Core network.* In both models we considered, kinetic proofreading reduces the absolute value of the error-length coefficient without changing its sign. To show this effect in general, we consider an arbitrary reaction network where we identify some of the reaction steps as those implementing kinetic proofreading (Fig.4.a). For example, in both models of Fig. 2, the proofreading reactions are those with rates $k_p^r$ and $k_p^w$. The complete network has an error probability $\eta_0$ and an error-length coefficient $r_{\eta L}$. We now remove all the proofreading reactions and define the remaining reactions as the "core network". In many practical case the core network is a simple linear chain of reactions, so that it is easy to compute its error probability $\eta_0^{\text{core}}$ and its error-length coefficient $r_{\eta L}^{\text{core}}$. To compare $r_{\eta L}$ and $r_{\eta L}^{\text{core}}$ we assume that both $\eta_0$ and $\eta_0^{\text{core}}$ are small so that Eq. (9) holds. We further assume that proofreading is a relatively rare event that does not significantly influence the incorporation times. Taking the ratio $r_{\eta L}/r_{\eta L}^{\text{core}}$ we therefore obtain

$$r_{\eta L} \approx \sqrt{\frac{\eta_0}{\eta_0^{\text{core}}}}\, r_{\eta L}^{\text{core}}. \qquad (14)$$

Since proofreading reduces the error probability ($\eta_0 < \eta_0^{\text{core}}$), it also reduces the absolute value of the error-length coefficient without changing its sign. We tested our prediction by computing $r_{\eta L}$ from experimentally measured kinetic rates in *E. coli* ribosomes (Table S1, SI, and [14, 19]) and from the T7 DNA polymerase [10] (see Figure S1 for the T7 datum). Eq. (14) qualitatively captures the dependence of the error-length coefficient on the error-correction effectiveness (Fig. 4.b). Quantitative discrepancies arise because the assumption that proofreading does not affect incorporation times partially breaks down.

The core-network approach qualitatively explains why the error-speed correlations have different signs in the Hopfield model (Figure 2.a) and in the protein translation model (Figure 2.b). Because of the discrimination in the backward rates, $r$ monomer bind to the enzyme for a long time in the core network of the Hopfield before the final incorporation. On the other hand, $w$ monomers bind to the enzyme for a short time before they are either rejected or incorporated. This implies that $r_{\eta L}^{\text{core}} > 0$ and therefore $r_{\eta L} > 0$, as showed in Figure 3. Conversely, the discrimination in the forward rates grants a fast incorporation of $r$ monomers in the core network of the protein translation model. Thus, $r_{\eta L}^{\text{core}} < 0$ and $r_{\eta L} < 0$, consistently with Eq. (13).

In this paper we studied the correlations of the empirical error in a copy polymer and its synthesis speed. These correlations probe general features of error-correction and permit to classify error-correction mechanisms into two broad categories: those leading to positive or to negative error-speed correlations. We showed that the Hopfield model [20] and a model of protein translation with discrimination in the forward rates [12–14, 26, 28] belong to opposite categories. Furthermore, a model of T7 DNA polymerase with forward discrimination [10] belongs to the same category of the protein translation model (see SI). This suggests that measurements of the error-speed correlations could reveal the presence of forward discrimination in replicative enzymes. Cell-free translation systems [29, 30] could provide simple and versatile in vitro assays to perform these measurements for ribosomes. A possible experiment would be to let the system translate for a fixed short time, separate the products into shorter and longer peptides, and then measure using mass spectroscopy whether the two categories contain significantly different errors. Similar experiments for DNA polymerases could bring insight into poorly characterized chemical reaction networks, such as those of human mitochondrial DNA pol-$\gamma$ [31], yeast pol-$\epsilon$ [32] and pol-$\delta$ [33, 34].

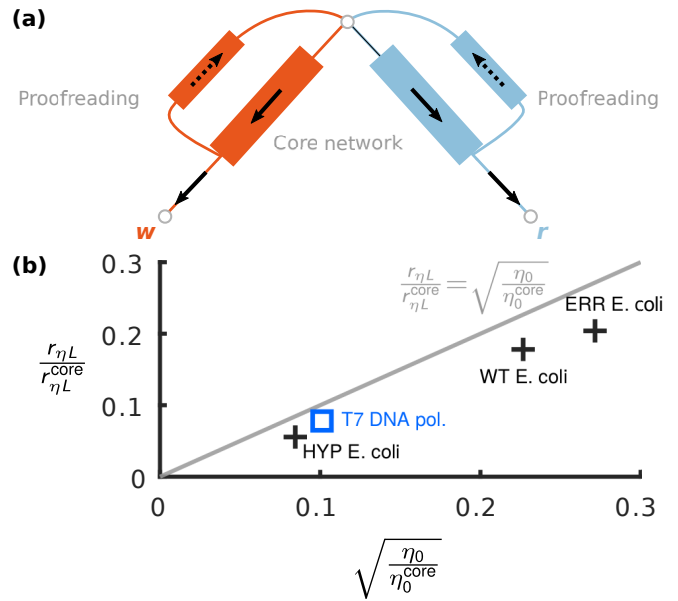

FIG. 4. Proofreading suppresses the error-speed correlations. (a) Incorporation with a "core network" complemented by proofreading reactions. Each block in the figure represents an arbitrary sub-network with an average flux in the direction of the arrows. (b) Comparison of Eq. (14) (solid line) with computation of error-speed correlations from measured kinetic rates, see SI. We considered the ribosomes in three strains of *E. coli*: wild type, hypercorrective, and error-prone [14, 19]. For each strain, we built the core network by removing the proofreading reactions and computed the relative change in $r_{\eta L}$ and $\eta_0$ between the original and core networks. We performed the same analysis for a model of T7 DNA polymerase (blue square, see SI and [8]). The data qualitatively agree with Eq. (14).

The magnitude of the error-speed correlations decreases when proofreading effectiveness increases. This implies that proofreading-deficient enzymes [31–34] and in-vitro assays that favor mis-incorporation [11, 28] are best suited to test our theory, for two reasons. First, the increased magnitude of error-speed correlations in the absence of error correction makes them easier to measure. Second, the poor precision of proofreading-deficient enzymes [10] reduces the sample size needed to empirically estimate error fluctuations.

Our result may also have consequences for the evolution of genomes. Recent work showed that the cells which replicates earlier thanks to environmental fluctuations, contributes more to population growth [35]. With significant error-speed correlations, the growth of a population could then be driven by the individuals whose DNA and proteins have significantly different error fractions from the population average. This phenomenon could have played a role in early stages of life.

We underline the conceptual difference between our results and the speed-error trade-off [5, 6, 19, 25] in particular as observed in protein translation [26, 28, 36]. In translation, tuning the concentration of $Mg^{++}$ ions provokes an approximately linear trade-off between the average error and the average reaction speed [26]. This kind of tradeoffs may depend on the choice of a control parameter [6, 19]. In contrast, we have shown that fluctuations of velocity and error are negatively correlated in protein translation for fixed external parameters. It remains to be explored whether the two results can be generally connected, in a similar fashion as equilibrium fluctuations and responses to external forces are related in statistical physics [37, 38].

We thank Michael Baym, Lucas Carey, Antonio Celani, Massimo Cencini, Todd Gingrich, and Jordan Horowitz for discussion. We further thank S. Aird and P. Laurino for comments on the manuscript. This work was supported by JSPS KAKENHI Grant Number JP18K03473 (to DC and SP).

# Supplementary Information for: Error-speed correlations in biopolymer synthesis

Davide Chiuchiú,[1] Yuhai Tu,[2] and Simone Pigolotti[1, *]

[1]*Biological Complexity Unit, Okinawa Institute of Science and Technology Graduate University, Onna, Okinawa 904-0495, Japan*
[2]*IBM T.J. Watson Research Center, Yorktown Heights, NY 10598, U.S.A.*

This document contains additional material supplementing the manuscript "Error-speed correlations in biopolymer replication" (from now on "Main Text").

The document is organized as follow. In Section I we prove the equivalence between fixed-length and fixed-time setups. In Section II we derive our main result on error-speed correlations, i.e. Eqs. (7)-(8) in the Main Text. In Section III we discuss how to compute moments of incorporation times distribution using first-passage time techniques. In Section IV we discuss numerical computations. In particular, we present simulations to validate Eq. (4), Eqs. (7)-(8) and present details for computation of data in Figures 4 and 3. Section V presents computation of the error rates for proofreading models. Section VI provides the reaction network for the T7 DNA polymerase together with the measured kinetic rates (Fig. SI.3). It also shows how the datum in Fig. 4 of the Main Text was computed. In Tables SI.2 and SI.3 we provide details on the distribution of the kinetic rates used in Fig. 3 of Main Text

## I. EQUIVALENCE BETWEEN FIXED-LENGTH AND FIXED-TIME SETUPS

To prove the equivalence of the fixed-length and fixed-time setups, we assume that the distribution of $\{\mathcal{L} = L/T, \mathcal{W} = W/T\}$ at fixed $T$, and that of $\{\mathcal{T} = T/L, \eta = W/L\}$ at fixed $L$ satisfy large deviation principles [9]

$$P(\mathcal{L}, \mathcal{W}|T) \asymp \exp[-TI(\mathcal{L}, \mathcal{W})] \tag{SI.1a}$$

$$P(\mathcal{T}, \eta|L) \asymp \exp[-L\phi(\mathcal{T}, \eta)] \tag{SI.1b}$$

where $\asymp$ indicates the leading behavior of the distributions for large $T$ and $L$. The rate functions $I$ and $\psi$ attain their minimum at the average values of $\{\mathcal{L}, \mathcal{W}\}$ and $\{\mathcal{T}, \eta\}$, respectively. Their Hessian matrices evaluated at the minimum are proportional to the inverse covariance matrices of $\{\mathcal{L}, \mathcal{W}\}$ and $\{\mathcal{T}, \eta\}$.

To connect $I$ and $\phi$ we use that, for large times $T$, the distribution of a given observable $X$ determines the time distribution to observe a fixed value of $X$ [4]. In particular,

$$\psi\left(\frac{T}{X}\right) = \frac{T}{X} J\left(\frac{X}{T}\right) \tag{SI.2}$$

where $\psi$ and $J$ are the rate functions of the intensive variables $T/X$ and $X/T$ for large values of $X$ and $T$ respectively. We assume that this result generalizes to joint distributions and apply it to $I$ and $\phi$. This yields

$$\phi(\mathcal{T}, \eta) = \mathcal{T} I\left(\frac{1}{\mathcal{T}}, \frac{\eta}{\mathcal{T}}\right). \tag{SI.3}$$

We now perform the change of variable $\mathcal{W} = \eta\mathcal{L}$ in Eq. (SI.1a) to obtain the joint probability $P(\mathcal{L}, \eta|T)$ of $\mathcal{L}$ and $\eta$ at a fixed time $T$. At the leading order $P(\mathcal{L}, \eta|T) \asymp \exp[-TQ(\mathcal{L}, \eta)]$, where

$$Q(\mathcal{L}, \eta) = I(\mathcal{L}, \eta\mathcal{L}). \tag{SI.4}$$

Combining Eq. (SI.3) and Eq. (SI.4) and expressing variances and covariances using the Hessian matrices of $I$, $\psi$ and $Q$, we obtain

$$r_{\eta\mathcal{T}} = -\left.\frac{\partial_{\mathcal{T}\eta}\phi}{\sqrt{\partial_{\mathcal{T}\mathcal{T}}\phi\,\partial_{\eta\eta}\phi}}\right|_{\min\phi} = \left.\frac{\partial_{\mathcal{L}\eta}Q}{\sqrt{\partial_{\mathcal{L}\mathcal{L}}Q\,\partial_{\eta\eta}Q}}\right|_{\min Q} = -r_{\eta\mathcal{L}}. \tag{SI.5}$$

which is equivalent to

$$r_{\eta T} = -r_{\eta L}, \tag{SI.6}$$

when passing to extensive variables. Equation (SI.6) has been validated using numerical simulations for different incorporation schemes. See Section IV for details.

---

[*]Electronic address: `simone.pigolotti@oist.jp`

## II. DERIVATION OF THE GENERAL FORMULA FOR THE ERROR-SPEED CORRELATIONS

The probability to have produced a copy polymer made up of a number $R$ of right monomers and a number $W$ of wrong ones at a given time $T$ can be approximated as

$$P(R,W|T) \approx \binom{R+W}{W} \eta_0^W (1-\eta_0)^R \int_0^\infty \delta\left(\sum_{n=1}^R \tau_n + \sum_{m=1}^W \tau_m - T\right) \left[\prod_{i=1}^R P(\tau_i|r)\mathrm{d}\tau_i\right]\left[\prod_{j=1}^W P(\tau_j|w)\mathrm{d}\tau_j\right]. \quad (\text{SI.7})$$

Here, the binomial term counts all the possible permutations of $R$ right monomers and $W$ wrong ones. The integral term with the Dirac delta function $\delta(\cdot)$ isolate the trajectories with the prescribed number of right and wrong monomers at time $T$. We used the approximation sign since Eq. (SI.7) implicitly assumes that the sum of incorporation times is exactly equal to $T$, whereas in practice it can be equal to $T' < T$ if there are no further incorporation in the time interval $[T',T]$. Representing the delta function as

$$\delta(x) = \int_{-\infty}^\infty \frac{e^{isx}}{2\pi} \mathrm{d}s, \quad (\text{SI.8})$$

and swapping the integration order we obtain

$$\rho(R,W,T) \sim \binom{R+W}{W} \eta_0^W (1-\eta_0)^R \int_{-\infty}^\infty \mathrm{d}s \, \exp\left[R\ln \tilde{P}(s|r) + W\ln \tilde{P}(s|w) - isT\right] \quad (\text{SI.9})$$

where

$$\tilde{P}(s|x) = \int_0^\infty \mathrm{d}\tau \, P(\tau|x) e^{is\tau} \quad (\text{SI.10})$$

is the cumulant generating function of $\tau$ conditioned to the incorporation of monomer $x$. We assume that both $P(\tau|r)$ and $P(\tau|w)$ have a finite mean and variance. Under this hypothesis, the central limit theorem ensures that the sum of random incorporation times in Eq.(SI.7) tends to a Gaussian random variable. This implies that we can truncate the cumulant generating function $\ln \tilde{P}(s|x)$ as

$$\ln \tilde{P}(s|x) \sim i\langle \tau \rangle_x s - \frac{\sigma_{\tau,x}^2 s^2}{2}. \quad (\text{SI.11})$$

Substituting (SI.11) into (SI.9), using the Stirling approximation, and omitting sub-dominant terms finally gives the expression

$$P(R,W|T) \approx \exp\left[-(R+W)D_{KL}^{\eta,\eta_0} - \frac{(\langle\tau\rangle_r R + \langle\tau\rangle_w W - T)^2}{2(R\sigma_{\tau,r}^2 + W\sigma_{\tau,w}^2)}\right] \quad (\text{SI.12})$$

where

$$D_{KL}^{\eta,\eta_0} = -(1-\eta)\ln[(1-\eta_0)(1-\eta)] - \eta\ln(\eta_0/\eta) \quad (\text{SI.13})$$

is the Kullback-Leibler divergence between the error probability $\eta_0$ and the error $\eta = W/(R+W)$. To compute $\sigma_R^2$, $\sigma_W^2$ and $\sigma_{RW}^2$ we approximate Eq.(SI.12) as a bivariate Gaussian distribution around Eq.(SI.12) maximum. This gives

$$\sigma_R^2 = C \, \frac{\eta_0 \langle\tau\rangle_w^2 + (1-\eta_0)\left((1-\eta_0)\sigma_{\tau,r}^2 + \eta_0 \sigma_{\tau,w}^2\right)}{\eta_0} \quad (\text{SI.14a})$$

$$\sigma_W^2 = C \, \frac{(1-\eta_0)\langle\tau\rangle_r^2 + \eta_0\left((1-\eta_0)\sigma_{\tau,r}^2 + \eta_0 \sigma_{\tau,w}^2\right)}{(1-\eta_0)} \quad (\text{SI.14b})$$

$$\sigma_{RW}^2 = C \left[(1-\eta_0)\sigma_{\tau,r}^2 + \eta_0 \sigma_{\tau,w}^2 - \langle\tau\rangle_r\langle\tau\rangle_w\right] \quad (\text{SI.14c})$$

$$\quad (\text{SI.14d})$$

where $C$ is a multiplicative factor. Substituting these expressions in Eq.(5) of the Main Text finally gives our main result, Eqs. (7)-(8). We also validated Eqs. (7)-(8) numerically, see Section IV.

## III. MOMENTS OF THE INCORPORATION TIMES

Analytical expressions for the first and second moments of the incorporation times are necessary to numerically evaluate Eqs. (7)-(8) in real cases. To derive such quantities, we treat monomer incorporation as a first-passage problem. We consider the probability $P_{x,i}(\tau)$ that incorporation takes a time $\tau$ given that the incorporated monomer is $x \in \{r, w\}$ and the initial state of the network is $i$. In the theory of stochastic processes, $P_{x,i}(\tau)$ represents the first-passage time distribution to reach the absorbing state $x$ from state $i$ [1, 2]. The Laplace transforms $\tilde{P}_{x,i}(s) = \int_0^\infty e^{-s\tau} P_{x,i}(\tau) d\tau$ evolve according to [1, 2, 8]

$$s\tilde{P}_{x,i}(s) = \sum_{j \notin \{i, r^*, w^*\}} k_{j,i} \left[\tilde{P}_{r,j}(s) - \tilde{P}_{r,i}(s)\right] + k_{x^*,i} - (k_{r^*,i} + k_{w^*,i})\tilde{P}_{r,i}(s) \tag{SI.15}$$

where $k_{j,i}$ is the rate of the reaction from state $i$ to $j$, '0' labels the network state where monomer incorporation starts, while $r^*$ and $w^*$ label the states where right and wrong monomers are finally incorporated, respectively. This means that conditional incorporation time distribution used in the Main Text is equal to $P(\tau|x) = P_{x,0}(\tau)$. From the solution of Eq. (SI.15) we obtain [1, 2, 8]

$$\eta_0 = \tilde{P}_{w,0}(0), \quad \langle\tau\rangle_x = -\frac{(\partial_s \tilde{P}_{x,0})(0)}{\tilde{P}_{x,0}(0)}, \quad \sigma^2_{\tau,x} = \frac{(\partial_{ss} \tilde{P}_{x,0})(0)}{\tilde{P}_{x,0}(0)} - \langle\tau\rangle^2_x. \tag{SI.16}$$

## IV. NUMERICAL COMPUTATIONS

To validate Eq. (4) we considered two different incorporation processes. In the first, incorporation requires two consecutive irreversible reactions (Network 1 in Figure SI.1.a). In the second process, incorporation follows the Michaelis-Menten enzyme dynamics (Network 2 in Figure SI.1.a). For both processes, we randomly and independently generated the reaction rates in the range $[0.01, 10]$ 36 times for each reaction network, and then computed $r_{\eta L}$ and $r_{\eta T}$ by averaging over 5000 trajectories of the polymerization process for each set of rates. Stochastic trajectories are simulated with the Gillespie algorithm [3]. For the setup at fixed length, we stopped each simulation when the polymer has reached a length $L = 5000$. In the setup at fixed time we run simulations until a time $T$ such that $\langle L \rangle = 5000$. Prediction given by Eq. (4) agree with the numerical simulations (see Figure SI.1.b).

To validate Eqs. (7)-(8) we compare the values of $r_{\eta L}$ from the numerical simulations in Figure SI.1.b with their corresponding predictions obtained by substituting Eqs. (SI.16) in Eqs. (7)-(8). Theoretical predictions agrees with the simulated data (see Figure SI.1.c), supporting the validity of Eqs. (7)-(8).

To generate the data points shown in Figures 3 and 4, we use Eq. (SI.16) and (7)-(8) to evaluate $\eta_0$ and $r_{\eta L}$ for the two reaction networks in Figure 2 with given kinetic rates. Data for the Hopfield model (Figure 4.a) are obtained from 60 independent random choices of the reaction rates (see Table SI.2 for the generation rules). We also generated 60 random sets of reaction rates for protein translation (black squares in Figure 4.b), see Table SI.3. In the same figure, blue crosses represent actual experimentally measured rates in different strains of E. coli [1]. The datum for the T7 DNA polymerase is obtained in the same way but using a different reaction network. See Section VI for details.

## V. ERROR-SPEED CORRELATIONS IN PROOFREADING MODELS

*Hopfield model:* We now apply Eq. (SI.15) and Eq. (SI.16) to the Hopfield model of Figure 2a. Combining the expressions for $\eta_0$, $\langle\tau\rangle_x$ and $\sigma^2_{\tau,x}$ together with Eq. (7) and Eq. (8), taking the limit $k \to \infty$, and expanding for small $n$ gives

$$\eta_0 \approx \left(1 + e^{\frac{2(\Delta E_w - \Delta E_r)}{k_B T}}\right)^{-1} \tag{SI.17a}$$

$$r_{\eta L} \approx \frac{\left(e^{\frac{\Delta E_w}{k_B T}} - e^{\frac{\Delta E_r}{k_B T}}\right)n}{e^{\frac{2\Delta E_w}{k_B T}}\left(1 + e^{\frac{\Delta E_r}{k_B T}}\right) + e^{\frac{2\Delta E_r}{k_B T}}\left(1 + e^{\frac{\Delta E_w}{k_B T}}\right) + e^{\frac{2(\Delta E_w + \Delta E_r)}{k_B T}}}. \tag{SI.17b}$$

To obtain Eq. (11) of the Main Text we observe that $r_{\eta L} \geq 0$ when $\Delta E_r \leq \Delta E_w$. Moreover $n \leq \exp[\Delta E_r/k_B T]$ because the probability to incorporate a $r$ monomer must be larger than the probability of proofreading it in useful operating regimes. Considering then the largest possible value for $\eta$ in Eq. (SI.17b), taking the limit $\Delta E_r \ll -1$ and then using Eq. (SI.17a) finally gives Eq. (11) of the Main Text.

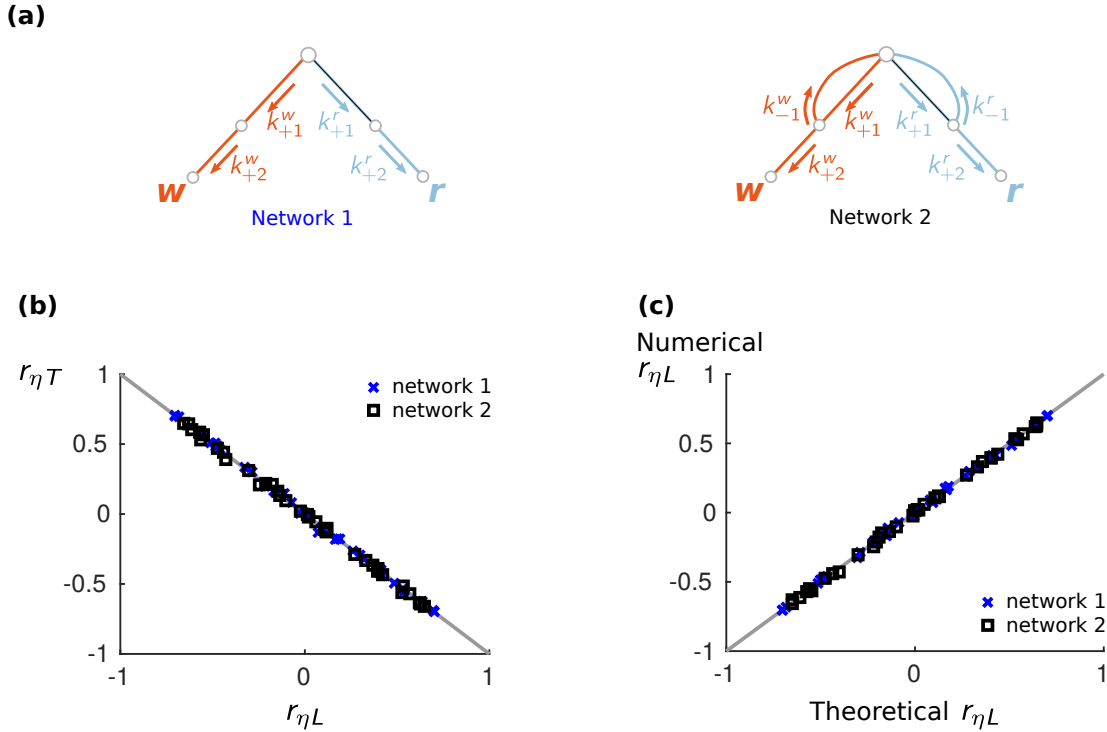

FIG. SI.1: Numerical tests of Eq. (4) and Eqs. (7)-(8). (a) Reaction networks for synthesis. Network 1 (left): monomer incorporation takes place after two consecutive irreversible reactions. Network 2 (right): monomer incorporation follows the Michaelis-Menten enzyme dynamics. (b) Equivalence of the fixed time and fixed length setups. Each point correspond to the numerical value of $r_{\eta L}$ and $r_{\eta T}$ obtained for an independent random choice of the reaction rates upon averaging over 5000 realizations of the polymerization process. For each realization we take $L = 5000$ as stopping conditions for the fixed-length setup, and $T$ such that $\langle L \rangle = 5000$ as the stopping condition in the fixed-time setup. Prediction given by Eq. (4) (grey line) agree with numerical simulations. (c) Validation of Eqs. (7)-(8). Each point compares the numerical values of $r_{\eta L}$ used to test Eq. (4) with the corresponding theoretical predictions obtained by substituting Eqs. (SI.16) into Eqs. (7)-(8) for each choice of the reaction rates.

*Protein translation model:* to apply Eq. (SI.15) and Eq. (SI.16) to protein translation, we consider the model of Figure 2a with rate values as in [1] (see Table SI.1). To reduce the number of parameters, we introduce a minimal model of protein translation in which the rates whose average over the three *E. coli* strands is of the same order of magnitudes in [1] are assigned the same value (see Figure SI.2 and Table SI.1). We apply Eq. (SI.15) and Eq. (SI.16) to such minimal model to compute $\eta_0$ and $r_{\eta L}$. Expanding these quantities for large $k_+$ gives

$$\eta_0 \approx \frac{k_\epsilon}{k_+}\left(1 + \frac{k_\epsilon}{k}\right)^{-1} \tag{SI.18a}$$

$$r_{\eta L} \approx -\frac{1}{\sqrt{2}}\left(1 + \frac{k}{k_\epsilon}\right)^{-\frac{1}{2}}. \tag{SI.18b}$$

These equations are equivalent to Eqs. (12) and (13) in the Main Text upon substituting again the rates $k_+$, $k$ $k_\epsilon$ of the minimal model with the specific rates they originate from.

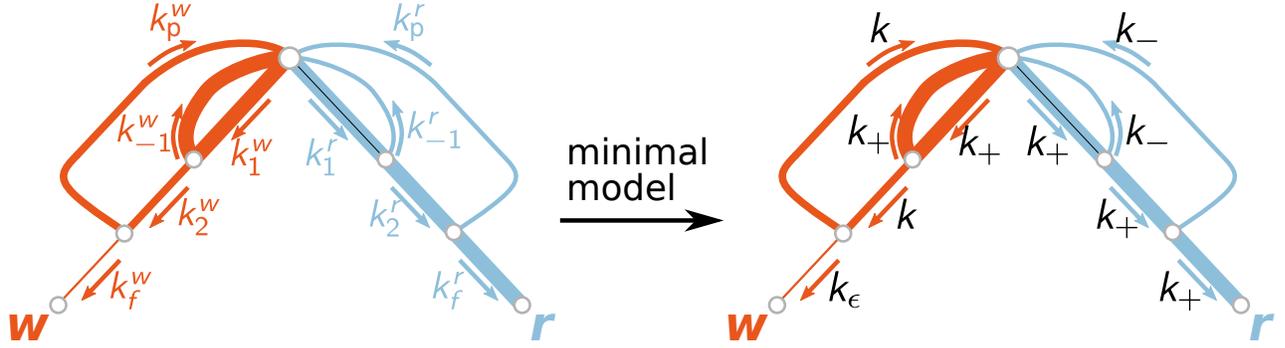

FIG. SI.2: Minimal model of protein translation where we consider identical all the reaction rates with same orders of magnitudes in the original model of protein translation.

| rate | WT [$s^{-1}$] | HYP [$s^{-1}$] | ERR [$s^{-1}$] | minimal model |
|---|---|---|---|---|
| $k_1^r$ | 40 | 27 | 37 | $k_+$ |
| $k_{-1}^r$ | 0.5 | 0.41 | 0.43 | $k_-$ |
| $k_2^r$ | 25 | 14 | 31 | $k_+$ |
| $k_p^r$ | $8.5 \times 10^{-2}$ | $4.8 \times 10^{-2}$ | $7.7 \times 10^{-2}$ | $k_-$ |
| $k_f^r$ | 8.415 | 4.752 | 7.623 | $k_+$ |
| $k_1^w$ | 27 | 25 | 36 | $k_+$ |
| $k_{-1}^w$ | 47 | 46 | 4 | $k_+$ |
| $k_2^w$ | 1.2 | 0.49 | 3.9 | $k$ |
| $k_p^w$ | 0.67 | 0.50 | 0.59 | $k$ |
| $k_f^w$ | $3.53 \times 10^{-2}$ | $3.52 \times 10^{-3}$ | $3.125 \times 10^{-2}$ | $k_\epsilon$ |

TABLE SI.1: Kinetic rates for the protein translation model measured in three different strands of *E. coli* from [1, 7]: wild type (WT), hypercorrective rpsL141 mutant (HYP) and error prone rps D12 mutant (ERR). In the column *minimal model*, we assign a single parameter to rates with the same order of magnitude. More specifically, we computed the average value of each rate for the three *E. coli* strands. We then assigned the value $k_\epsilon$ if the average is less than $0.05 s^{-1}$; $k_-$ if the average is in the range $[0.05, 0.5]$ $s^{-1}$; $k$ if the average is in $[0.5, 5]$ $s^{-1}$; and $k_+$ if the average is greater than $5 s^{-1}$.

## VI. T7 DNA POLYMERASE

We consider the reaction network in Figure SI.3 to model the incorporation process by the T7 DNA polymerase. This network correspond to the one in [5] where we neglected polymerase detachment. The rates are $k^r_{\text{pol}}$ =300 s$^{-1}$, $k^r_{\text{pp}}$ =100 s$^{-1}$, $k^r_{\text{next}}$ =300 s$^{-1}$, $k^r_{\text{sx}}$ =0.2 s$^{-1}$, $k^r_{\text{sp}}$ =700 s$^{-1}$, $k^w_{\text{pol}}$ =0.03 s$^{-1}$, $k^w_{\text{pp}}$ <1 × 10$^{-5}$ s$^{-1}$, $k^w_{\text{next}}$ = 0.01 s$^{-1}$, $k^w_{\text{sx}}$ =2.3 s$^{-1}$, $k^w_{\text{sp}}$ =700 s$^{-1}$ and $k^w_{\text{exo}}$ =900 s$^{-1}$. To obtain the datum shown in Figure 4 of the Main Text, we first computed $\eta_0$ and $r_{\eta L}$ for the full network with the help of Eqs. (7)-(8) of the main text and Eqs. (SI.15)-(SI.16). Substituting the rates for T7 DNA polymerases gives $\eta_0 = 1 \times 10^{-6}$ and $r_{\eta L} = -0.06$. We then repeated the procedure for the core network defined upon removing the proofreading reaction (Figure SI.3), and obtained $\eta_0^{\text{core}} = 1 \times 10^{-4}$ and $r_{\eta L}^{\text{core}} = -0.71$. Note that $r_{\eta L} < 0$ in both the original and the core network.

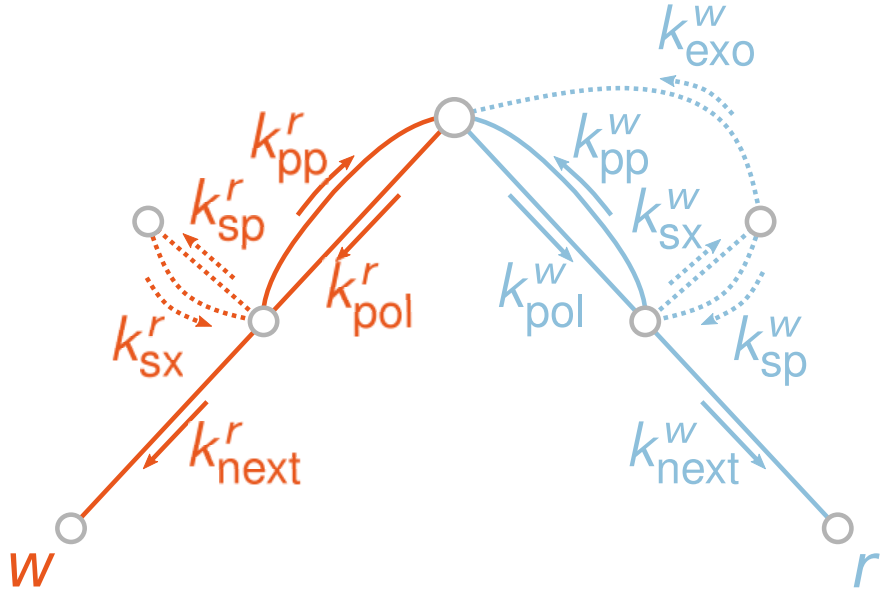

FIG. SI.3: Reaction network of the T7 DNA polymease. Dashed lines represent reactions that are removed from the complete network to obtain the core network used in Figure 4.b.

**Distribution of rates for Fig. 3 of the Main Text**

| Quantity | Uniform in |
|---|---|
| $\log_{10}(k)$ | $[2; 6]$ |
| $\log_{10}(-\Delta E_w/k_B T)$ | $[-1, 6]$ |
| $\log_{10}((\Delta E_w - \Delta E_r)/k_B T)$ | $[0, 4]$ |
| $\log_{10}(n) - \Delta E_r/k_B T$ | $[-4.5, 0]$ |

TABLE SI.2: Distribution of the random rates for the data in Figure 3.a of the Main Text. In this range of rates, the model always operates in the error-correction regime [6].

| Quantity | Uniform in |
|---|---|
| $k_3^w$ | 1 |
| $\log_{10}(k_1^r)$ | $[2, 4]$ |
| $\log_{10}(k_2^r/k_1^r)$ | $[-1, 1]$ |
| $\log_{10}(k_f^r/k_1^r)$ | $[-1, 1]$ |
| $\log_{10}(k_1^w/k_1^r)$ | $[-1, 1]$ |
| $\log_{10}(k_{-1}^w/k_1^r)$ | $[-1, 1]$ |
| $k_2^w$ | $[0.5, 1.5]$ |
| $k_{-1}^r$ | $[0.05, 0.5]$ |
| $k_r^3$ | $[0.05, 0.5]$ |
| $\log_{10}(k_f^w)$ | $[-6, -1]$ |

TABLE SI.3: Distribution of the random rates for the data in Figure 3.a of the Main Text. Rates generated in these ranges are always consistent with the minimal model of Figure SI.2.